\documentclass[prb,aps,twocolumn,superscriptaddress]{revtex4-1}
\usepackage{amssymb}
\usepackage{color}
\usepackage{graphicx}
\usepackage{dcolumn}
\usepackage{bm}
\usepackage[header,title,page,titletoc]{appendix}
\usepackage{amsmath}
\usepackage{amsfonts}%
\usepackage[colorlinks,linkcolor=red,anchorcolor=green,citecolor=blue]{hyperref}
\providecommand{\U}[1]{\protect \rule{.1in}{.1in}}

\begin{document}

\title{Electronic Bloch oscillation in a pristine monolayer graphene}

\author{Tongyun Huang}
\affiliation{Department of Physics, Beijing Normal University, Beijing
100875, China}
\affiliation{Beijing Computational Science Research Center, Beijing
100193, China}

\author{Ruofan Chen}
\affiliation{Beijing Computational Science Research Center, Beijing
100193, China}

\author{Tianxing Ma}
\email{txma@bnu.edu.cn}
\affiliation{Department of Physics, Beijing Normal University, Beijing
100875, China}
\affiliation{Beijing Computational Science Research Center, Beijing
100193, China}
\author{Li-Gang Wang}
\email{sxwlg@yahoo.com}
\affiliation{Zhejiang University, Hangzhou 310027, China}
\author{Hai-Qing Lin}
\affiliation{Beijing Computational Science Research Center, Beijing
100193, China}

\begin{abstract}
In a pristine monolayer graphene subjected to a constant electric field along the layer, the Bloch oscillation of an electron is studied in a simple and efficient way. By using the electronic dispersion relation, the formula of a semi-classical velocity is derived analytically, and then many aspects of Bloch oscillation,
such as its frequency, amplitude, as well as the direction of the oscillation, are investigated.
It is interesting to find that the electric field affects the component of motion, which is non-collinear with electric field,
and leads the particle to be accelerated or oscillated in another component.
\end{abstract}

\maketitle




%
In the solid state physics, Bloch oscillation is an important phenomenon. It is usually
involved with the coherent motion of quantum particles in periodic
structures. For example, an electron (a matter wave) suffers this
effect in a periodic lattice subjected to a constant external field.
This phenomenon is predicted from quantum mechanics in very early
days\cite{bloch,zener} and has been demonstrated in various fields of physics,
such as semiconductor superlattices\cite{Waschke,Abumo}, photonic
crystals\cite{optical}, cold-atom systems\cite{Dahan1996}, and
acoustic waves\cite{Sanchis2007}. However, electric domains lead to the instability of the electric field and destroy the Bloch oscillation in the semiconductor superlattices, it requires a complex design to suppress electric domains\cite{Feil2005}.

On the other hand, since its discovery in 2004, graphene has attracted a
tremendous amount of interest due to its unique properties that may
promise a broad range of potential applications\cite{Novoselov2004,Novoselov2005,neto,Ma2010,Peng2011,Raccichini2015}. Recently, many
theoretical and experimental investigations focus on the
graphene-based superlattices with electrostatic potentials or
magnetic barriers\cite{Bai2007,Park2008,Barbier2008,Jannik2008,Ramezani2008,Dell2009,Cheng2014,Lu2015,Mishra2017}, including periodic\cite{Hwan2008,Guo2011,Guinea2008,Wang2010},
aperiodic\cite{Zhao2011,Liang2012,Zhang2012},disorder\cite{Zhao2012}, and sheet arrays system\cite{Fan2014}. Different from the common semiconductors, graphene superlattices can maintain a stable electric field due to the uniform population of the quantum well, which is induced by the back gate voltage, and some researchers have investigated the electronic Bloch oscillations in a
structure with periodic potentials\cite{Dragoman2008}, a graphene
nanoribbon with a hybird superlattice\cite{Diaz2014}, graphene superlattices with multiple Zener tunneling\cite{Krueckl2012}, a tilted honeycomb lattice for the localized Wannier-Stark states\cite{ Kolovsky2013}, as well as the Bloch scillations in the gapped graphene. It has been demonstrated that Bloch oscillaions in graphene are different than in common semiconductors, since the electron in graphene is described by Dirac rather than the Schr\"{o}dinger equation. Furthermore, the Bloch oscillation in graphene superlattices has potential applications such as infrared detectors and lasers, One important issue still remains, that is, what does the electronic Bloch
oscillation behavior in the gapless graphene? 

Many aspects of Bloch oscillation
can be obtained by a single band description via using the dispersion
relation to derive the semi-classical velocity of the particle. In
this work, based on the electronic structure
under tight-binding approximation, we derive the motion of an electron
in pristine monolayer graphene subjected to a constant external field. Within such a simple and efficient way, our results show several
interesting phenomena of the electronic Bloch oscillation in graphene. For example, when the electric field is applied in one direction, the oscillation disappears in the x direction in a special condition, while it never happen in the y direction. Due to the linear dispersion relation, the amplitude and period of the oscillation are doubled as the particle passes through Dirac points, and its trajectory is almost a circle. 
In the following, we firstly derive the general formula of the motion of an electron based on the
dispersion relations, and then we analyze the properties of the
Bloch oscillation.

A monolayer graphene is well known for its honeycomb
structure, and its dispersion relation can be written as\cite{neto}
\begin{equation}
{\cal E}(k)=\pm \varepsilon\sqrt{3+f(k)},\label{EnergyBand}
\end{equation}
where
$f(k)=2\cos(\sqrt{3}ak_y)+4\cos(\frac{\sqrt{3}}{2}ak_y)
\cos(\frac{3}{2}ak_x)$.
$a\approx1.42{\AA}$
is the carbon-carbon distance, and 
$\varepsilon\approx2.5{\rm eV}$ is related to Fermi velocity ($v_F\approx10^6\rm\,m/s$), $\hbar
v_F=\frac{3}{2}\varepsilon a$\cite{neto}. The signs ``$+$'' and ``$-$'' are,
respectively, corresponding to the electron and hole energy band, which touch
together at Dirac points (DPs). From Eq. \ref{EnergyBand}, it is easy to find that the DPs are located at
$[\frac{4n\pi}{3a},\frac{2}{\sqrt{3}a}(2n\pm\frac{2}{3})\pi],
[\frac{2}{3a}(2n+1)\pi,\frac{2}{\sqrt{3}a}(2n\pm\frac{1}{3})\pi]$
with $n=0,\pm1,\pm2,\cdots$. According to $v=\frac{1}{\hbar}\frac{\partial{\cal
    E}(k)}{\partial k}$\cite{solid}, we can readily have
$v=(v_x,v_y)$ as the function of $k_x$ and $k_y$,
\begin{eqnarray}
v_x&=&\frac{\mp 3\varepsilon a\sin(\frac{3}{2}ak_x)\cos(\frac{\sqrt{3}}{2}ak_y)}{\hbar\sqrt{3+f(k)}}, \notag \\
v_y&=&\frac{\mp\sqrt{3} \varepsilon  a[\sin(\sqrt{3}ak_y)+\sin(\frac{\sqrt{3}}{2}ak_y)\cos(\frac{3}{2}ak_x)]}{\hbar\sqrt{3+f(k)}},
\end{eqnarray}
which show that $v_x$ and $v_y$ are the periodic functions of $k_x$ and $k_y$, 
and the sign ``$-$'' (``$+$'') is corresponding to the velocities of
electron ( hole or hole-like electron). Basically,  the Berry curvature could affect the trajectory of a wave packet undergoing Bloch oscillations in optical lattice\cite{Price2012}, twhile in present system there is no anomalous contribution, as the Berry curvature is just a monopole like contribution at the Dirac point for gapless graphene\cite{Xiao2010}. When a constant electric field $E=(E_x,E_y)$ is applied along the layer of graphene, D\'{o}ra et al. show that the velocity of massless Dirac electrons is pinned to the Fermi velocity in a finite field, and the electric field moves the Dirac point around in momentum space. Those special features imply that Dirac electrons in the electric field can be treated as critical particles, their motion is a drift transport, so they move ballistically and leave their footprints \cite{Dóra2010}. Thus, the semiclassical approach is valid, and we employ the electronic motion equation $\hbar\frac{dk(t)}{dt}=-eE$ to describe the motion of Dirac electron, which survives as 
\begin{equation}
 k_x(t)=k_x(0)-\frac{eE_x}{\hbar}t,\quad
 k_y(t)=k_y(0)-\frac{eE_y}{\hbar}t, \label{k-t}
\end{equation}
where $k_x(0)$ and $k_y(0)$ are the initial wave-vector values.
Substituting Eq. \ref{k-t} into Eq. 2, we can obtain the
dynamic formula for $v$. Therefore using Eqs. 2-3, we can analyze the motion of the electron or hole in graphene under the constant electric field. In particular, from Eq. (1), one can see that tight-binding approximations could describe both the conduction band and valence band, and the Dirac point moves continuously in momentum space and have not been destroyed under an electric field. On the other hand, previous study has also found that the behavior of the electron obtained by tight-binding is consistent with that from Bloch equations as the electron passes the Dirac point\cite{Hartmann2004,Ishikawa2010}. Thus, our formula is valid for dynamics involving band-crossing points. 
Since the direction of the electric field $E$ can be chosen
arbitrarily, we shall firstly discuss the electronic motion
when $E$ is only along the $x$ (\textit{case I }) or $y$ (\textit{case II}) direction and then generalize it to an arbitrary direction (\textit{case III}).


\textit{Case I: $E$ along the $x$ direction}. In
this case $E_y=0$, so we have
$k_y(t)=k_y={\rm constant}$, and $k_x(t)=k_x(0)-\frac{eE_x}{\hbar}t$.
Assuming $k_x(0)=0$, the dynamic formula of $v_{x}$ and $v_{y}$ are
\begin{subequations}
\label{E:decomposition}
\begin{eqnarray}
\label{eq:T1}
 v_x(t)&=&\frac{\mp3\varepsilon a\sin(-\frac{2\pi}{T}t)\cos(\frac{\sqrt{3}}{2}ak_y)}{\hbar G(t)},  \notag \\
 \end{eqnarray}
 \begin{eqnarray}
\label{eq:J1}
 v_y(t)&=&\frac{\mp\sqrt{3}\varepsilon a[\sin(\sqrt{3}ak_y)+\sin(\frac{\sqrt{3}}{2}ak_y)\cos(-\frac{2\pi}{T}t)]}{\hbar G(t)},
\end{eqnarray}
\end{subequations}
where $G(t)$=$\sqrt{3+2\cos(\sqrt{3}ak_y)+4\cos(-\frac{2\pi}{T}t)\cos(\frac{\sqrt{3}}{2}ak_y)}$,
and $T=\frac{4\pi}{3}\frac{\hbar}{|aeE_x|}$. From Eqs. 4(a)-(b), it is
easy to see that $v(t+T)=v(t)$ with $T$ being
the period of the motion. The frequency and circular frequency of
the Bloch oscillation are generally given by
\begin{equation}
\nu_B=\frac{1}{T}=\frac{3}{4\pi}\frac{|aeE_x|}{\hbar}\hbox{ and }
\omega_B=2\pi\nu_B=\frac{3}{2}\frac{|aeE_x|}{\hbar},
\end{equation}
respectively. According to the expression of $v$, the time-dependent position $r(t)$ of the electron is
$r(t)=r(0)+\int_0^t vdt$, and here we assume the initial position
$r(0)=0$, i.e., $x(0)=0$ and $y(0)=0$. After a simple derivation,
we obtain
$x(t)=C-\frac{s}{eE_x}G(t)$
where $C$ is an integration constant satisfying $x(0)=0$. For $y(t)$, we have
to numerically calculate the following
\begin{eqnarray}
y(t)&=&-\frac{2\sqrt{3}\varepsilon\omega_B}{3eE_x} \notag \\
&\times&\int_0^t
\frac{\sin(\sqrt{3}ak_y)+\sin(\frac{\sqrt{3}}{2}ak_y)
\cos(-\omega_Bt)}{G(t)}dt.
\end{eqnarray}
 According to the formula of $x(t)$, we can have
\begin{eqnarray}
x_{\rm max}=&&C-\frac{\varepsilon}{|eE_x|}\sqrt{3+2\cos(\sqrt{3}ak_y)-4|\cos(\frac{\sqrt{3}}{2}ak_y)|},\notag \\
x_{\rm min}=&&C-\frac{\varepsilon}{|eE_x|}\sqrt{3+2\cos(\sqrt{3}ak_y)+4|\cos(\frac{\sqrt{3}}{2}ak_y)|}, \notag \\
\end{eqnarray}
Therefore, the amplitude of the oscillation along $x$ direction,
$L_x=|x_{\rm max}-x_{\rm min}|$, is given by
\begin{equation}
L_x=\frac{\varepsilon}{|eE_x|}||1+2\cos(\frac{\sqrt{3}}{2}ak_y)|
-|1-2\cos(\frac{\sqrt{3}}{2}ak_y)||.\label{Lx}
\end{equation}
When
$|\cos(\frac{\sqrt{3}}{2}k_ya)|\ge\frac{1}{2}$,
$L_x$ has its maximum value: $L_x^{\rm max}=\frac{2\varepsilon}{|eE_x|}$.
When $\frac{\sqrt{3}}{2}ak_y=(n+\frac{1}{2})\pi$, $L_x^{\rm min}=0$. According to Eq. \ref{Lx}, if $E_x=4.61\,{\rm
  mV/nm}$, ( we set $\varepsilon=2.5\,{\rm eV}$ in the whole paper ), $L_x^{\rm
  max}\approx1084\,{\rm nm}$ with $\nu_B\approx237 {\rm
  GHz}$. The amplitude and period of the clean graphene is larger than those of superlattices based on the graphene with the gapped band structure, which are around 30nm and 0.8ps, respectively\cite{Krueckl2012}.

\begin{figure}[tbp]
\includegraphics[scale=0.45]{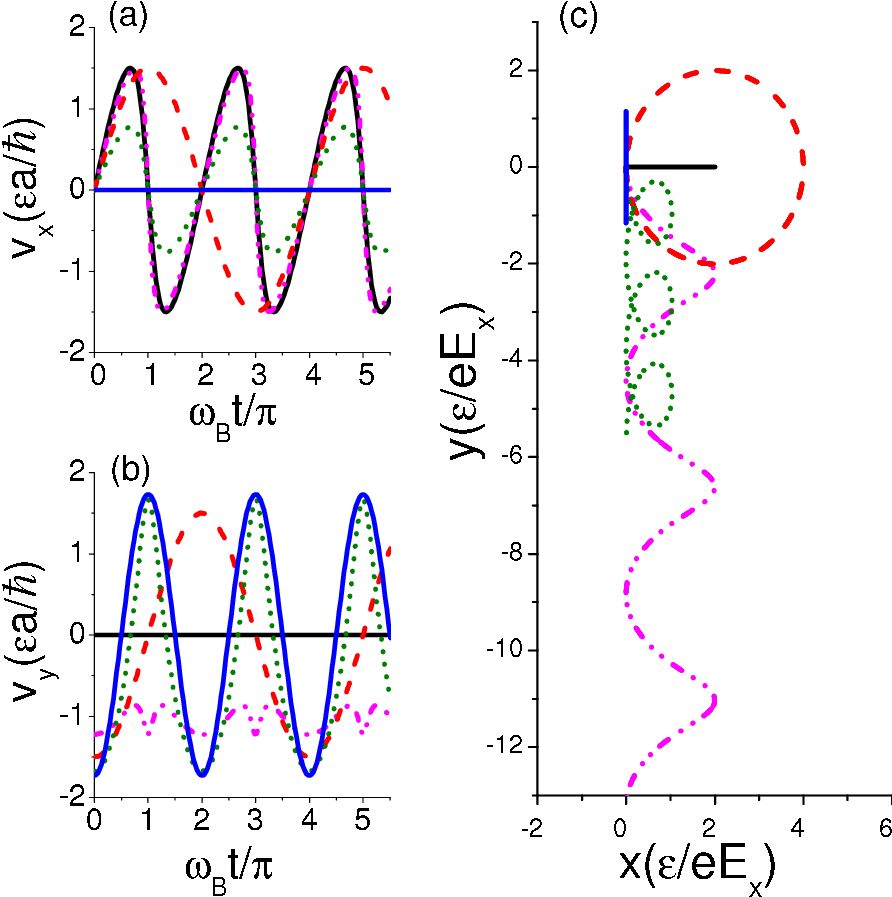}
\caption{(Color online) The time dependence of $v_x$ (a), $v_y$ (b) and (c) the trajectories of the electron (or hole-like electron)
on the graphene sheet for different $k_y$ at $\frac{\sqrt{3}}{2}ak_y=0$ ( solid dark line line ),
$\frac{\sqrt{3}}{2}ak_y=\frac{\pi}{4}$ (dash-dot-dot magenta line),
$\frac{\sqrt{3}}{2}ak_y=\frac{\pi}{3}$ (dash red line),
$\frac{\sqrt{3}}{2}ak_y=\frac{5}{12}\pi$ (dot green line), and
$\frac{\sqrt{3}}{2}ak_y=\frac{\pi}{2}$ (solid blue line). The electric field $\bm{E}$ is
along the $x$ direction.
}
\label{Fig:TM}
\end{figure}

Fig. 1(a) and (b) demonstrate the time dependence of $v_x$ and $v_y$ with
different values of $k_y$. It is found that when
$\frac{\sqrt{3}}{2}ak_y=\frac{\pi}{3}$, shown as the dash red lines, the electron passes through
the DPs, and the period of $v_x$ and $v_y$ is doubled. Because
the electron passes through the DPs, the electron
transits into another band and  behaves as a hole-like electron. 
After a period in another band, the hole-like electron behaves as the electron again. Therefore the period of the velocity
becomes twice time, and correspondingly the amplitude is also doubled.
This is quite different from the gapped case, where the Bloch oscillations originate from the interference between the electron and hole states\cite{Krueckl2012}. 
There is an interesting phenomenon that, the oscillation along the
$x$ direction disappears although $E$ is still along the $x$
direction when
$\frac{\sqrt{3}}{2}ak_y=(n+\frac{1}{2})\pi$. Meanwhile, since $v_y\ne0$, the oscillation in the $y$
direction remains, see 
the solid blue lines ($\frac{\sqrt{3}}{2}ak_y=\frac{\pi}{2}$) in
Fig. 1(a) and (b). In other hands, if $\frac{\sqrt{3}}{2}ak_y=n\pi$,
we have $v_y=0$. It means that the oscillation in $y$-axis
disappears and the oscillation in $x$-axis remains, see the solid dark lines in Fig. 1(a) and (b).

The corresponding electron's trajectories are shown in Fig. 1(c), where we demonstrate the trajectory of an electron within
three periods on the graphene layer. It is clear that, when the
electron (or hole) passes through the DPs, its amplitude is
doubled and its trajectory is approximately a circle, see the dash red lines ($\frac{\sqrt{3}}{2}ak_y=\frac{\pi}{3}$) in Fig. 1(c). The same phenomenon has also been found in graphene with an ultrashort intense terahertz radiation pulse\cite{Hartmann2004}. 
In general, the motion in the $x$ and $y$ directions may oscillate, but its
trajectory is very complex and depends on the initial value of $k_y$.
For example, its trajectory is a helix (see short dash green line for
$\frac{\sqrt{3}}{2}ak_y=\frac{5}{12}\pi$); or it may go further and further
with variational velocity, its trajectory is like a sine
function (see dash-dot-dot magenta line for $\frac{\sqrt{3}}{2}ak_y=\frac{\pi}{4}$).

\textit{Case II: $E$ along the $y$ direction}. In
this case, $E_x=0$, so we have
$k_x(t)=k_x={\rm constant}$ and $k_y(t)=k_y(0)-\frac{eE_y}{\hbar}t$.
Assuming $k_y(0)=0$, the dynamic formula read as
%
\begin{eqnarray}
v_x(t)&=&\frac{\mp3\varepsilon a\sin\frac{3}{2}(ak_x)
\cos(-\frac{2\pi}{T'}t)}{\hbar G'(t)},  \notag \\
v_y(t)&=&\frac{\mp\sqrt{3}\varepsilon a[\sin(\frac{-4\pi}{T'}t)+\sin(-\frac{2\pi}{T'}t)
\cos(\frac{3}{2}ak_x)]}{\hbar G'(t)}  \label{V2},
\end{eqnarray}
where $G'(t)=\sqrt{3+2\cos(-\frac{4\pi}{T'}t)+
4\cos(\frac{3}{2}ak_x)\cos(-\frac{2\pi}{T'}t)}$ and the period of the
motion $T'=\frac{4\pi}{3}\frac{\hbar}{|aeE_y|}$.
From Eq. \ref{V2}, we still have $v(t+T')=v(t)$. The frequency and circular frequency of Bloch oscillation are
accordingly
\begin{equation}
\nu'_B=\frac{1}{T'}=\frac{\sqrt{3}}{4\pi}\frac{|aeE_y|}{\hbar} \hbox{ and }
\omega'_B=\frac{\sqrt{3}}{2}\frac{|aeE_y|}{\hbar}.
\end{equation}
From Eqs. 5 and 9, we can see that the direction of the electric
field affects the frequency of Bloch oscillation.
Similar to \textit{Case I},
$y(t)=C'-\frac{\varepsilon}{eE_y}G'(t)$,
where $C'$ is an integration constant satisfying $y(0)=0$.
For $x(t)$ we have to numerically calculate
\begin{equation}
x(t)=-\frac{2\sqrt{3}\varepsilon\omega'_B}{eE_y}\int_0^t\frac{
\sin\left(\frac{3}{2}ak_x\right)\cos\left(-\omega'_Bt\right)}{G'(t)}dt.
\end{equation}
The amplitude of the oscillation
along the $y$ direction, $L_y=|y_{\rm max}-y_{\rm min}|$, is given by
\begin{equation}
L_y=\frac{\varepsilon}{|eE_y|}|\sqrt{5+
|4\cos(\frac{3}{2}ak_x)|}
-\sqrt{1-\cos^2(\frac{3}{2}ak_x)}\;|.
\end{equation}
It shows that $L_y$ has its maximum value $L_y^{\rm
  max}=\frac{3\varepsilon}{|eE_y|}$ when $\cos(\frac{3}{2}ak_x)=\pm1$,
and $L_y^{\rm min}=\frac{(\sqrt{5}-1)\varepsilon}{|eE_y|}$ when
$\cos(\frac{3}{2}ak_x)=0$. If $E_y=4.61\,{\rm mV/nm}$, then
$L_y^{\max}\approx1627\,{\rm nm}$, $L_y^{\rm min}\approx671\,{\rm
  nm}$ with  $\nu'_B\approx137\,{\rm GHz}$.

\begin{figure}[tbp]
\includegraphics[scale=0.45]{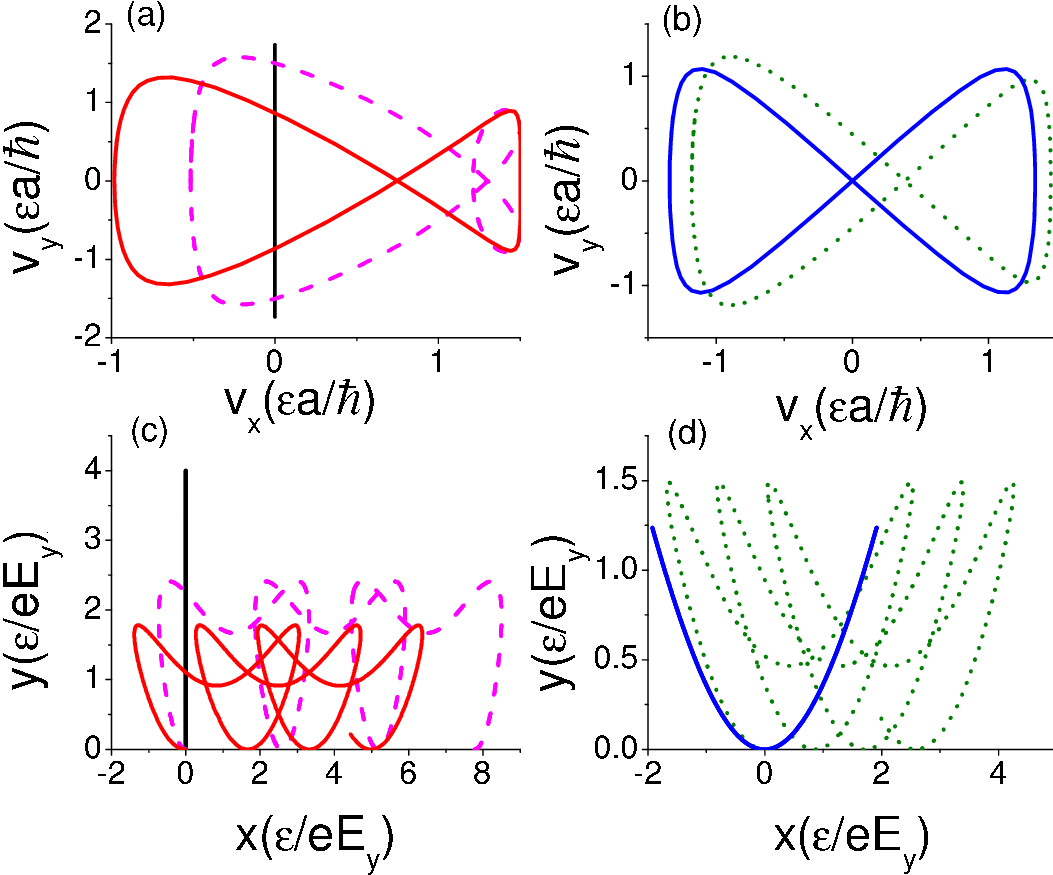}
\caption{(Color online) The Lissajous figures of $v_x$ (a) and $v_y$ (b) for different
values of $k_x$, and the trajectories of the electron (or hole-like
eletron) on the graphene sheet are shown in (c) and (d) with
$\frac{\sqrt{3}}{2}ak_x=0$ (solid dark line),
$\frac{\sqrt{3}}{2}ak_x=\frac{\pi}{6}$ (dash magenta line),
$\frac{\sqrt{3}}{2}ak_x=\frac{\pi}{3}$ (solid red line),
$\frac{\sqrt{3}}{2}ak_x=\frac{5}{12}\pi$ (dot green line), and
$\frac{\sqrt{3}}{2}ak_x=\frac{\pi}{2}$ (solid blue line). The electric field $E$ is
along the $y$ direction.}
\label{Fig:TM2}
\end{figure}
Fig. 2(a) and (b) show the Lissajous figures of $v_x$ and $v_y$ with
different values of $k_x$. When $\frac{\sqrt{3}}{2}ak_x=0$, the
electron passes through the DPs and behaves as a hole-like
electron, and after a period of time it behaves as an electron again,
and its amplitude increases accordingly. At this case when
$-\omega't\in[2n+\frac{2}{3}\pi, 2n+\frac{4}{3}\pi]$, the particle behaves as
a hole-like electron, otherwise it behaves as an electron.

Different from the \textit{case I}, the oscillation along the $y$
direction never disappears when electric field is along the $y$
direction, and its amplitude never becomes zero. There is a special case
that when $k_x=0$, we find that $v_x=0$. At this case the oscillation in
the $x$-direction may disappear. The corresponding trajectories within three
periods are shown in Fig. 2(c) and (d).
%

\textit{Case III: $E$ along the arbitrary direction.}
At this case, the dynamics of the electron becomes much more
complicated, since both $E_x$ and $E_y$ are non-zero. Meanwhile, the
dynamic properties of $v_x(t)$ and $v_y(t)$ are also related with the
initial phase $k_x(0)$ and $k_y(0)$ and the ratio $E_x/E_y$.

$v_x$ and $v_y$ depend on the two periodic functions $\cos\left(\frac{3}{2}a[k_x(0)-\frac{eE_x}{\hbar}t]\right)$ and
$\cos(\frac{\sqrt{3}}{2}a[k_y(0)-\frac{eE_y}{\hbar}t])$.
Let $T_x$ denote the period of the former, and $T_y$ denote the period of
the latter, so we have $T_x=\frac{4\pi}{3}\frac{\hbar}{|aeE_x|}$ and $T_y=\frac{4\pi}{\sqrt{3}}\frac{\hbar}{|aeE_y|}$. If the ratio
$T_x/T_y$ is rational, i.e. $T_x/T_y=m/n$, ($n$ and $m$ are integers),
then $v_x(t)$ and $v_y(t)$ are periodic with the periods being $mT_y$ or
$nT_x$. But if $T_x/T_y$ is irrational, $v_x(t)$ and $v_y(t)$ are not
periodic anymore: they have not a finite period. In this case
the motion is not a periodic oscillation, even though the particle may still move back and forth. Thus, regular Bloch oscillations merge in the rational direction of the electric field, this semi-classical result is consistent with results in the previous work, which is obtained by quantum theory\cite{ Kolovsky2013}. 

\begin{figure}[tbp]
\includegraphics[scale=0.45]{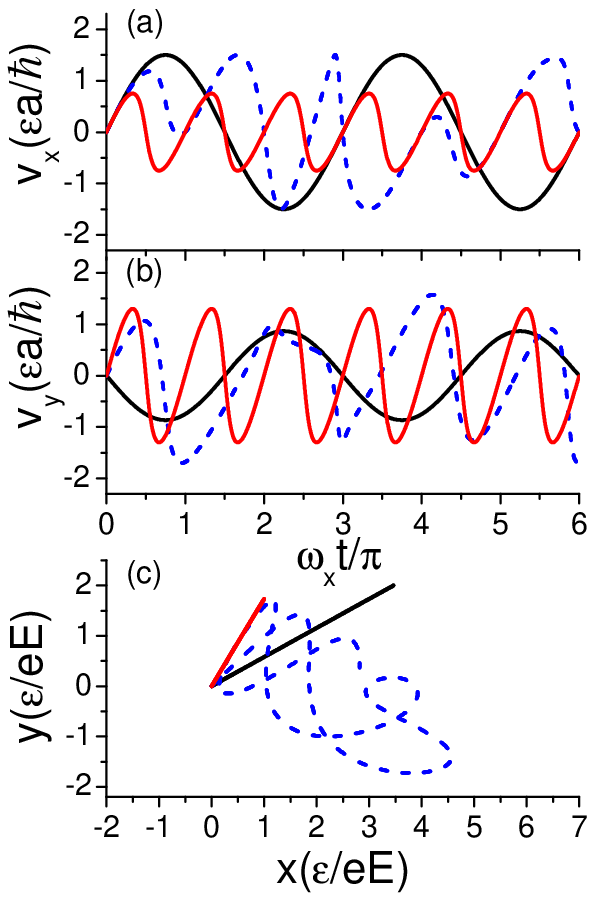}
\caption{(Color online) The time dependence of (a) $v_x$,
(b) $v_y$ and (c) the trajectories of the electron (or hole-like electron) on the graphene sheet for $\alpha=\frac{\pi}{6}$(solid
dark line), $\alpha=\frac{\pi}{4}$ (dash blue line) and $\alpha=\frac{\pi}{3}$ (solid red line) . }
\label{Fig:TM3}
\end{figure}
Finally, we discuss the dynamics of electrons under the
condition of $k_x(0)=k_y(0)=0$ with some specific directions, where
$x(0)=y(0)=0$, $\alpha=\arctan(E_y/E_x)$, $E=|\bm{E}|$, and $\omega_x=2\pi/T_x$. The time dependence of $v_x$ and $v_y$ with different
values of $\alpha$ is illustrated in Fig. 3(a) and (b), and 
the corresponding trajectories within $3T_x$ are shown in Fig.
3 (c). Due to the rotational symmetrical structure of graphene, it is
easy to find that, when $\alpha=\frac{\pi}{6}$ (solid dark line) the direction of the
electric field is equivalent to the $y$ direction, and the electron
also passes through DPs in this case; when
$\alpha=\frac{\pi}{3}$ (solid red line) the direction of the electric field is
equivalent to the $x$ direction. When
$\frac{\sqrt{3}}{2}ak_y=\frac{\pi}{4}$ (dash blue line), it is a general case, in which
the electron only moves within a single energy band.
%
%

In summary, we have derived the general formulas for the velocity of the
electron in graphene subject to a constant electric field, and have analyzed the dynamic properties of electron for some 
particular and interesting 
cases.
When electric field is along $x$-axis and $y$-axis, we find Bloch oscillation in direction of electric
field, and we obtain formulas for its amplitude and frequency. We
also find that the electric field affects the motion in other
direction, making the electron oscillating or moving forward with
fluctuation in other direction. Moreover, the
velocity is periodic in all directions in these two cases. Finally, we analyze the
period of the motion and present the numerical result if electric field
has an arbitrary direction. Our result provides a positive insight for
experimentally observing the Bloch oscillation in a pristine graphene, which may facilitate the development of graphene-based electronics.

This work is supported by NSFCs (Grant. No. 11774033 and 11674284), Fundamental Research Funds for the Center Universities (No.
2017FZA3005), National Key Research and Development Program of China (No. 2017YFA0304202), and Zhejiang provincial Nature science Foundation of China (No. LD18A040001). We also acknowledge the support from by the HSCC of Beijing Normal University, and the Special Program for Applied Research on Super Computation of the NSFC-Guangdong Joint Fund (the second phase).


\begin{thebibliography}{99}\setlength{\itemsep}{0pt}
\bibitem{bloch} F.~Bloch, Z.~Phys. {\bf 52}, 555 (1929).
\bibitem{zener}C.~Zener, Proc.~R.~Soc.~London A {\bf 145}, 523 (1934).
\bibitem{Waschke}C.~Waschke, H.~G.~Roskos, R~Schwedler, K.~Leo, H~Kurz,
and K.~K\"ohler, Phys.~Rev.~Lett. {\bf 70}, 3319 (1993).
\bibitem{Abumo}P.~Abumov, and D.~W.~L.~Sprung, Phys.~Rev.~B {\bf 75}, 165421
(2007).
\bibitem{optical}R.~Sapienza, P.~Costantino, and D.~Wiersma
Phys.~Rev.~Lett {\bf 91}, 263902 (2003).
\bibitem{Dahan1996}M.~B.~Dahan, E.~Peik, J.~Peichel, Y.~Castin, and
C.~Salomon, Phys.~Rev.~Lett. {\bf 76}, 4508 (1996).
\bibitem{Sanchis2007}H.~Sanchis-Alepuz, Y.~Kosevich, and J.~S\'anchez-Dehesa,
Phys.~Rev.~Lett. {\bf 98}, 134301 (2007).
\bibitem{Feil2005}T. Feil, H.-P. Tranitz, M. Reinwald, and W. Wegscheider, Appl.~Phys.~Lett. {\bf 87}, 212112 (2005).
\bibitem{Novoselov2004}K.~S.~Novoselov, A.~K.~Geim, S.~V.~Morozov,
  D.~Jiang, Y.~Zhang, S.~V.~Dubonos, I.~V.~Grigorieva, and
  A.~A.~Firsov, Science {\bf 306}, 666(2004).
\bibitem{Novoselov2005} K.~S.~Novoselov, A.~K.~Geim, S.~V.~Morozov,
D.~Jiang, M.~I.~Katsnelson, I.~V.~Grigorieva1,
S.~V.~Dubonos, and A.~A.~Firsov, Nature {\bf 438}, 197-200 (2005).
\bibitem{neto}A.~H.~Castro~Neto, N.~M.~R.~Peres, K.~S.~Novoselov, and
  A.~K.~Geim, Rev.~Mod.~Phys. {\bf 81}, 109 (2009).
\bibitem{Ma2010} T. Ma, F. Hu, Z. Huang, and H.-Q. Lin,
Appl. Phys. Lett. \textbf{97}, 112504 (2010); F. Hu, T. Ma,
H.-Q Lin, and J. E. Gubernatis, Phys. Rev. B \textbf{84}, 075414 (2011).
\bibitem{Peng2011} X.-H. Peng, and S. Velasquez, Appl. Phys. Lett. {\bf 98}, 023112 (2011).
\bibitem{Raccichini2015}R.~Raccichini, A.~Varzi, S.~Passerini, and B.~Scrosati, Nature~Mater. {\bf14}, 271 (2015).
\bibitem{Bai2007}C.~Bai, and X.~Zhang, Phys.~Rev.~B {\bf 76}, 075430 (2007).
\bibitem{Park2008}C.~Park, Li Yang, Young-Woo Son, M.~L.~Cohen, and S.~ G.~Louie, Nat. Phys. {\bf 4}, 213 (2008).
\bibitem{Barbier2008} M.~Barbier, F.~M.~Peeters, P.~Vasilopoulos,
and J.~Milton Pereira, Phys.~Rev.~B. {\bf 77}, 115446 (2008).
\bibitem{Jannik2008}J.~C.~Meyer, C.~O.~Girit, M.~F.~Crommie, and
A.~Zettl, Appl.~Phys.~Lett. {\bf 92}, 123110 (2008).
\bibitem{Ramezani2008} M.~R.~Masir, P.~Vasilopoulos, A.~Matulis,
and F.~M.~Peeters, Phys.~Rev.~B {\bf 77}, 235443 (2008).
\bibitem{Dell2009} L.~D.~Anna, and A.~D.~Martino, Phys.~Rev.~B
{\bf 79}, 045420 (2009).
\bibitem{Cheng2014}H.~Cheng, C.~Li, T.~Ma, L.-G.~Wang, Y.~Song, and H-Q.~Lin, Appl.~Phys.~Lett. {\bf105}, 072103 (2014).
\bibitem{Lu2015}W.-T.~Lu, and W.~Li, Appl.~Phys.~Lett. {\bf107}, 082110 (2015).
\bibitem{Mishra2017}S.~K.~Mishra, A.~Kumar, C.~P.~Kaushik, and B.~Dikshit, J.~Appl.~Phys. {\bf121}, 184301 (2017).
\bibitem{Hwan2008} C.~Park, Li Yang, Young-Woo Son,
M.~L.~Cohen, and Steven G.~Louie, Phys.~Rev.~Lett. {\bf101} 126804 (2008).
\bibitem{Guo2011} Xiao-Xiao~Guo, De~Liu, and Yu-Xian Li,
  Appl.~Phys.~Lett. {\bf 98}, 242101 (2011).
\bibitem{Guinea2008} F.~Guinea, M.~I.~Katsnelson, and M.~A.~H. Vozmediano,
Phys.~Rev.~B {\bf 77}, 075422 (2008).
\bibitem{Wang2010}L.-G.~Wang, and S.-Y.~ Zhu,
Phys.~Rev.~B. {\bf 81}, 205444 (2010).
\bibitem{Zhao2011}P.-L.~Zhao, and X.~Chen, Appl.~Phys.~Lett. {\bf 98}, 242101 (2011).
\bibitem{Liang2012}T.~Ma, C.~Liang, L.-G.~Wang, and H.-Q.~Lin,
  Appl.~Phys.~Lett. {\bf 100}, 252402 (2012).
 \bibitem{Zhang2012} Z. Zhang, H. Li, Z. Gong, Y. Fan, T. Zhang, and H. Chen, Appl.~Phys.~Lett. {\bf 101}, 252104 (2012).
\bibitem{Zhao2012}Q.~Zhao, and J.~Gong, Phys.~Rev.~B. {\bf 85},
104201 (2012).
\bibitem{Fan2014}Y.~Fan, B.~Wang, H.~Huang, K.~Wang, H.~Long, and P.~Lu, Opt.~Lett. {\bf39}, 6827 (2014).
\bibitem{Dragoman2008}D.~Dragoman, and M.~Dragoman, Appl.~Phys.~Lett.
{\bf 93}, 103105 (2008).
\bibitem{Diaz2014}E. D\'{\i}az, K. Miralles, F. Dom\'{\i}nguez-Adame, and C. Gaul, Appl.~Phys.~Lett.
{\bf 105}, 103109 (2014).
\bibitem{Krueckl2012}Viktor Krueckl, and Klaus Richter, Phys.~Rev.~B {\bf 85}, 115433 (2012).
\bibitem{Kolovsky2013}Andrey R. Kolovsky, and Evgeny N. Bulgakov, Phys.~Rev.~A {\bf 87}, 033602 (2013).
\bibitem{solid} J.~M.~Ziman, Principles of the Theory of Solids (Cambridge University Press, 1972).
\bibitem{Price2012} H. M. Price, and N. R. Cooper, Phys.~Rev.~A {\bf85}, 033620 (2012).
\bibitem{Xiao2010} Di Xiao, Ming-Che Chang, and Qian Niu, Rev.~Mod.~Phys. {\bf82}, 1959 (2010).
\bibitem{Dóra2010} Bal\'{a}zs D\'{o}ra, and Roderich Moessner, Phys.~Rev.~B {\bf81}, 165431 (2010).
\bibitem{Hartmann2004} T. Hartmann, F. Keck, H. J. Korsch, and S. Mossmann, New J. Phys. {\bf6}, 2 (2004).
\bibitem{Ishikawa2010} Kenichi L. Ishikawa, Phys.~Rev.~B {\bf82}, 201402(R) (2010).






\end{thebibliography}
\end{document}